\RequirePackage[hyphens]{url}

\documentclass[10pt,conference]{IEEEtran}

\usepackage{amsmath,amssymb,amsfonts}
\usepackage{subcaption}
\usepackage{bbm,epsfig,bm}
\usepackage{algorithmic}
\usepackage{graphicx}
\usepackage{textcomp}
\usepackage{dcolumn}
\usepackage[hidelinks]{hyperref}
\usepackage{xspace}
\usepackage[normalem]{ulem}
\usepackage[binary-units=true]{siunitx}
\usepackage{booktabs}
\usepackage{arydshln} 
\def\BibTeX{{\rm B\kern-.05em{\sc i\kern-.025em b}\kern-.08em
    T\kern-.1667em\lower.7ex\hbox{E}\kern-.125emX}}
\usepackage[capitalise]{cleveref}
\usepackage{tikz}
\usetikzlibrary{calc,shapes,arrows,positioning}
\usepackage{tcolorbox}
\usepackage{balance}
\usepackage{wrapfig}
\usepackage{framed}
\usepackage{caption}
\usepackage[T1]{fontenc}
\usepackage{threeparttable}

\usepackage[backend=bibtex,style=ieee,maxcitenames=2,mincitenames=1,natbib=true]{biblatex} 
\newcounter{HCounter}

\usepackage{array}
\newcolumntype{P}[1]{>{\raggedright\arraybackslash}p{#1}}

\usepackage{array}
\newcommand{\PreserveBackslash}[1]{\let\temp=\\#1\let\\=\temp}
\newcolumntype{C}[1]{>{\PreserveBackslash\centering}p{#1}}
\newcolumntype{R}[1]{>{\PreserveBackslash\raggedleft}p{#1}}
\newcolumntype{L}[1]{>{\PreserveBackslash\raggedright}p{#1}}

\def\BibTeX{{\rm B\kern-.05em{\sc i\kern-.025em b}\kern-.08em
    T\kern-.1667em\lower.7ex\hbox{E}\kern-.125emX}}

\newcommand{\linebreakand}{%
  \end{@IEEEauthorhalign}
  \hfill\mbox{}\par
  \mbox{}\hfill\begin{@IEEEauthorhalign}
}

\usepackage{listings}
\definecolor{codegreen}{rgb}{0,0.6,0}
\definecolor{codegray}{rgb}{0.5,0.5,0.5}
\definecolor{codepurple}{rgb}{0.58,0,0.82}
\definecolor{backcolour}{rgb}{0.95,0.95,0.92}

\lstdefinestyle{mystyle}{
    backgroundcolor=\color{backcolour},   
    commentstyle=\color{codegreen},
    keywordstyle=\color{magenta},
    numberstyle=\tiny\color{codegray},
    stringstyle=\color{codepurple},
    basicstyle=\ttfamily\footnotesize,
    breakatwhitespace=false,         
    breaklines=true,                 
    captionpos=b,                    
    keepspaces=true,                 
    numbers=left,                    
    numbersep=5pt,                  
    showspaces=false,                
    showstringspaces=false,
    showtabs=false,                  
    tabsize=2
}
\title{The Dynamics of Innovation in Open Source Software Ecosystems}

\author{
    \IEEEauthorblockN{ G\'abor M\'esz\'aros\IEEEauthorrefmark{1} and Johannes Wachs\IEEEauthorrefmark{2}\IEEEauthorrefmark{3}\IEEEauthorrefmark{4}}
    \\
     \IEEEauthorblockA{\IEEEauthorrefmark{1}Center for Collective Learning, Corvinus University of Budapest, Hungary}
    \IEEEauthorblockA{\IEEEauthorrefmark{2}Department of Network Science, Corvinus University of Budapest, Hungary}
        \IEEEauthorblockA{\IEEEauthorrefmark{3}HUN-REN Centre for Economics and Regional Studies, Budapest, Hungary}
            \IEEEauthorblockA{\IEEEauthorrefmark{4}Complexity Science Hub, Vienna, Austria\\ johannes.wachs@uni-corvinus.hu}

}
\begin{document}






\maketitle

\begin{abstract}

Software libraries are the elementary building blocks of open source software ecosystems, extending the capabilities of programming languages beyond their standard libraries. Although ecosystem health is often quantified using data on libraries and their interdependencies, we know little about the rate at which new libraries are developed and used. Here we study imports of libraries in 12 different programming language ecosystems within millions of Stack Overflow posts over a 15 year period. New libraries emerge at a remarkably predictable sub-linear rate within ecosystems per post. As a consequence, the distribution of the frequency of use of libraries in all ecosystems is highly concentrated: the most widely used libraries are used many times more often than the average. Although new libraries come out more slowly over time, novel combinations of libraries appear at an approximately linear rate, suggesting that recombination is a key innovation process in software. Newer users are more likely to use new libraries and new combinations, and we find significant variation in the rates of innovation between countries. Our work links the evolution of OSS ecosystems to the literature on the dynamics of innovation, revealing how ecosystems grow and highlighting implications for sustainability.

\end{abstract}


\section{Introduction}

Software plays a key role in the modern digital economy \cite{andreessen2011software}, reflected in the explosive growth of software-related patents \cite{chattergoon2022winner} and the increasing importance of software in traditional sectors like the automobile industry \cite{charette2021car}. But how is the software industry itself evolving? Where do innovations in software come from? Empirically quantifying innovation in software is especially challenging \cite{bin2020impact}, in part because the usual proxies used in the innovation literature like patents \cite{mansfield1986patents} or trademarks \cite{castaldi2018trademark} are ill-suited to the software context. For instance patents tend to focus on end products and are broadly categorized; it also takes years from invention to award. They also ignore, by definition, open source software.

This is an important blindspot because of the crucial role that OSS plays in software development \cite{musseau2022open}, and by extension on the digital economy; OSS ecosystems are referred to as digital infrastructure in analogy to roads and bridges \cite{eghbal2016roads,eghbal2020working}. In particular, OSS libraries are like building blocks that developers use and combine to build projects \cite{fang2024novelty}. These blocks are codified, packaged, and made re-useable through social coding platforms like GitHub \cite{dabbish2012social}, creating large and growing ecosystems of interdependent code \cite{mens2014studying,mens2023software}. The importance of these OSS building blocks in the broader economy will only grow as more products integrate digital components \cite{hoffmann2024value,korkmaz2024github,gortmaker2024open,stojkoski2024estimating,juhasz2024software}. At the same time, researchers have raised concerns about the sustainability of these ecosystems because of their interdependencies and significant reliance on voluntary activity \cite{avelino2019abandonment,geiger2021labor,hejderup2022use,miller2023we}.

Indeed, these ecosystems thrive on the collaborative development and sharing of software libraries that developers can incorporate into their projects \cite{basili1996reuse}. However, our understanding of the dynamics governing the introduction of new libraries and their novel combinations is surprisingly limited. This gap in knowledge hinders our ability to foster innovation effectively, allocate resources wisely, and sustain the health of OSS ecosystems. Despite recent work \cite{fang2024novelty} on how developers combine libraries within projects, less is known about the overall rate of innovation at the ecosystem level. Understanding these rates is vital, as they shape the growth of dependency structures and reveal how libraries become central or peripheral to the broader ecosystem \cite{decan2019empirical}. 

In this paper we examine the dynamics of innovation in OSS ecosystems through a quantitative study of OSS library use in Stack Overflow posts. By extracting libraries used in millions of posts written in 12 different programming languages across 15 years of data, we can observe the rates which new libraries emerge in different ecosystems. We borrow concepts from the Schumpeterian school, which frames innovation as an evolutionary process which can be meaningfully studied by observing the dynamics of elementary ingredients used in a system \cite{schumpeter1939business,youn2015invention}. In our framing, OSS libraries are these ingredients. A key insight from this literature, already recognized in the software engineering research community \cite{fang2024novelty}, is the key role that combinations of elementary ingredients play in the generation of impactful innovation \cite{kogut1992knowledge,weitzman1998recombinant,fleming2001recombinant,uzzi2013atypical,youn2015invention}.

We first show that across all ecosystems new libraries appear at slowing rate, suggesting that innovation within ecosystems may slow down over time. A consequence is that over time, the distribution of library use becomes highly concentrated: for example, the 10\% most frequently imported Python libraries account for about 80\% of imports. On the other hand, the rate of novel pairs of libraries imported in posts grows at a linear rate in all ecosystems, suggesting that combinatorial innovation is the driver of growing ecosystems. This finding is similar to a study of over 200 years of novel codes and pairs of codes appearing on US patents \cite{youn2015invention}, which the authors interpret as demonstrating a constant rate of exploration versus exploitation in innovation over the long run. 

Finally we present results on the origin of innovation in software ecosystems. First we show that new users are significantly more likely to use new libraries and combinations, across all ecosystems studied. Second, after geolocating posting users, we report results on which countries are most likely to import new libraries or combinations. The geography of innovation in OSS ecosystems is surprisingly diverse.

Our results suggest that there are characteristic patterns of innovation in OSS ecosystems, and that these patterns can inform our understanding of ecosystem sustainability. New libraries emerge more slowly as ecosystems grow, concentrating use and hence risk in relatively few libraries. This also that maintenance of libraries can be prioritized. However, the steady growth of novel combinations of libraries used in code suggests that the need for maintenance to address issues emerging between ecosystem libraries (for instance through dependency links \cite{decan2017empirical,pashchenko2018vulnerable,zimmermann2019small,schueller2024modeling} or co-use relationships) will only grow. Finally, the observation that new and geographically diverse contributors are drivers of innovation in OSS ecosystems suggests the importance of dismantling barriers to OSS participation \cite{steinmacher2015social,mendez2018open} and facilitating onboarding \cite{trinkenreich2020hidden} for the sake of long run ecosystem health.

\section{Theoretical Framework}
We first review prior work on the dynamics of innovation, both generally and within software. We then examine how different issues in OSS maintenance relate to growing systems and their usage.

\subsection*{Dynamics of Innovation}
Innovation systems are often analyzed by examining their elementary ingredients—fundamental units that, when combined, generate ideas, products, or technologies. For example, recent work in economic development draws an analogy between a country's economic capabilities and the game of Scrabble, where the more letters (capabilities) a player gathers, the more high-value words (products) they can create \cite{hausmann2014atlas}.

A substantial body of research thus studies the rate at which new ingredients are introduced. Across a variety of domains, new ingredients appear at ever more slowly (i.e. sub-linear) rates. In linguistics the rate at which new words appear in a text decreases as the text grows longer \cite{baayen2012word}. Within individual software programs, program vocabulary grows sub-linearly with program length \cite{zhang2009discovering}. This pattern is known as Heaps' Law \cite{heaps1978information}, and has been observed in the emergence of online social annotations (i.e. tagging systems) \cite{cattuto2009collective}, Wikipedia pages \cite{tria2014dynamics}, and cooking recipes \cite{zhu2013geography}.

One important consequence of Heaps' Law is that the slowing rate of new ingredients in a system will naturally lead to a significant concentration of usage around specific ingredients \cite{tria2018zipf}. Such distributions follow power laws or Zipf's law \cite{newman2005power}. In such distributions, a small number of elements are highly prevalent while the majority are infrequently used, reflecting a significant concentration of activity around certain key components. In other words, in growing systems with a slowly growing set of ingredients, some ingredients will be used many times more often than others. Such power laws appear at various levels of abstraction in software: in links between Java classes, dependencies between Perl libraries, and FreeBSD Ports, among others \cite{valverde2002scale,louridas2008power}.

A second, more abstract consequence of Heaps' Law is that it suggests that either all such systems will tend to slow down over time, or that innovation is about more than just the emergence of new ingredients. Indeed, the literature since Schumpeter \cite{schumpeter1939business} has emphasized the importance of combinatorial innovation—the novel recombination of existing elements to create new functionalities or concepts. Indeed, if one considers new ways how ingredients can be combined, possibilities for innovation grow exponentially even if new ingredients emerge slowly \cite{weitzman1998recombinant}. The striking difference in the rates of introduction between single new components and novel combinations over time was confirmed through an analysis of 200 years of US patent data, which observed that while the rate of introducing new technological classes (individual novelties) slows down, patents registered with novel pairs of classes continue to appear at a linear rate \cite{youn2015invention}.

Empirical studies across various fields support the importance of combinatorial innovation. In technology and patent research, firms innovate by recombining their existing knowledge and capabilities, resulting in new products and processes \cite{kogut1992knowledge}. Analyses of patent data demonstrate that inventions stemming from novel combinations of existing technologies are more likely to be impactful, although they come with higher uncertainty \cite{fleming2001recombinant}. In scientific research, high-impact papers often arise from atypical combinations of established ideas, suggesting that blending familiar concepts in novel ways can lead to significant breakthroughs \cite{uzzi2013atypical}. 

What about software? Software is widely thought to be a highly innovative field \cite{chattergoon2022winner}. Researchers interested in the sustainability of software systems have increasingly focused on open source libraries as the building blocks or infrastructure of software \cite{eghbal2016roads}. This suggests that OSS libraries are the ingredients of software--like the words that makes up novels, technology classes assigned to patents, or the literal ingredients used to cook a recipe-- thus the right lens through which to study innovation in software. Indeed, open-source software itself in particular is considered an intensely creative activity \cite{lakhani2005hackers}, and reuse is a core concept in software engineering \cite{basili1996reuse,frakes2005software}. 

In fact, previous work has explored innovation within individual OSS projects by examining how the adoption of novel combinations of libraries influences project success \cite{fang2024novelty}. The authors suggest an innovation-maintenance trade-off: using novel combinations of libraries enhances a project's functionality and appeal but such combinations are more likely to be realized by smaller teams, which may be less able to maintain the project in the long run. Innovation may also the burden of maintenance due to added complexity and dependency management. Unlike patents, where maintenance requirements are minimal after the invention is secured, software demands continuous updates and support to remain functional and secure. Despite this emerging tension between innovation and sustainability in software, we do not know much about the dynamics of innovation at the ecosystem level.

\subsection*{Sustainability of growing OSS ecosystems}
OSS ecosystems are growing, both in terms of the number of libraries and the dependencies between them \cite{decan2019empirical}. Implications for maintainability and sustainability are well-studied: issues arising because of vulnerabilities \cite{zimmermann2019small,decan2018impact}, library abandonment \cite{miller2023we}, or update incompatibilities \cite{decan2017empirical} spread beyond individual libraries through dependencies to affect larger systems. 

At the same time, a better understanding of the dynamics of innovation in OSS ecosystems can inform our understanding of and barriers to their sustainability. For example, if OSS libraries appear at a sub-linear rate, it suggests that, over time, use of specific libraries will follow a highly skewed distribution. This implies that maintenance on a few specific libraries will be essential for sustainability. On the other hand, if growth in combinatorial innovation is linear, it suggests that cross-functionality of libraries is an important part of ecosystem health. Libraries used together are not necessarily near each other in the dependency tree, thus present a new perspective on links within an ecosystem. 

On the other hand, innovation itself is likely an important, if overlooked aspect of ecosystem health \cite{mens2023software}. Knowing about the rates of innovation in an OSS ecosystem, whether in terms of libraries or their combinations, can tell us if they go through different stages of maturity. If innovation is indeed important for ecosystem health, then it makes sense to know who is responsible for innovation. For instance, young people or people new to a scientific field are thought to be less beholden to established patterns, and thus be more able to innovate \cite{dietrich2007optimal}; confirmation that new users, or that users from different geographic locations, are more likely to generate innovations in the OSS context would provide a new motivation for supporting developer onboarding and support for new contributors \cite{gerosa2021shifting,trinkenreich2020hidden,steinmacher2015social}.

\section{Methods}

\subsection*{Data Source}

Our primary data source for this study is Stack Overflow, the leading Q\&A platform for programmers. We accessed the complete Stack Overflow data dump, which includes all posts from the platform's inception in 2008 up to 2024. This dataset is widely utilized in research to explore various aspects of software development and developer behavior \cite{anderson2012discovering, vasilescu2013stackoverflow, vasilescu2014social, xu2020makes}. A wide variety of people, from beginner to  experts are active on the platform \cite{vadlamani2020studying}. Indeed, Stack Overflow plays a significant role in the software development community by influencing coding practices \cite{vasilescu2013stackoverflow,wu2019developers} and even acting as a labor market signal \cite{xu2020makes}. Unlike traditional mailing lists or forums, Stack Overflow offers features such as collective curation and gamification, which enhance the quality and relevance of its content \cite{vasilescu2014social}.

Stack Overflow is well-suited for studying innovation dynamics in OSS ecosystems for several reasons. First, its posts typically represent solutions or attempts at solutions to specific programming problems, offering practical insights into real-world challenges faced by developers. Code snippets are direct attempts to solve these problems. Unlike public OSS repos like GitHub, data mined from Stack Overflow partially reflects what developers working on closed-sourced projects are doing. Second, Stack Overflow is highly curated: duplicate and off-topic posts are systematically removed, bot activity is minimized, and community editors actively refine content. Users are encouraged to include detailed explanations, code snippets, and examples, enhancing the clarity and utility of shared knowledge. This rigorous curation ensures that the data reflects genuine developer activity. Note that we do not claim that the Stack Overflow posts that first mention a specific library are the actual moment of innovation. Rather these events are signals of innovative activity within an ecosystem.

The data dump contains over 50 million posts, including more than 20 million questions and 30 million answers, contributed by approximately 17 million registered users. For our analysis, we focus on posts related to the 12 programming languages with the highest number of posts on the platform during our study period. These languages are: Python, R, JavaScript, Java, C++, PHP, Ruby, Perl, Rust, Swift, Objective-C, and C\#. This selection includes newer languages that were launched within our dataset's timeline, such as Rust and Swift, as well as a language like Objective-C, which has reached planned obsolescence. 

\subsection*{Data Extraction}
We assign posts to languages using tags, with answers inheriting the tags of their questions. To extract code snippets from these posts, we adapt code by Baltes et al.\cite{baltes2019sotorrent}. As a post may be tagged with multiple programming languages, it is possible that an individual post is scanned for language-specific libraries multiple times.

Given a set of code snippets within a post, we extract libraries using language-specific regular expressions. If a post contains code in another language (for example if a post is tagged with multiple languages), it is unlikely that the regular expression would extract any libraries. We only retain posts that have at least one import statement. We show an example question from Stack Overflow, tagged with Python, and including a code snippet with import statements.

\begin{figure}
    \centering
    \includegraphics[width=0.49\textwidth]{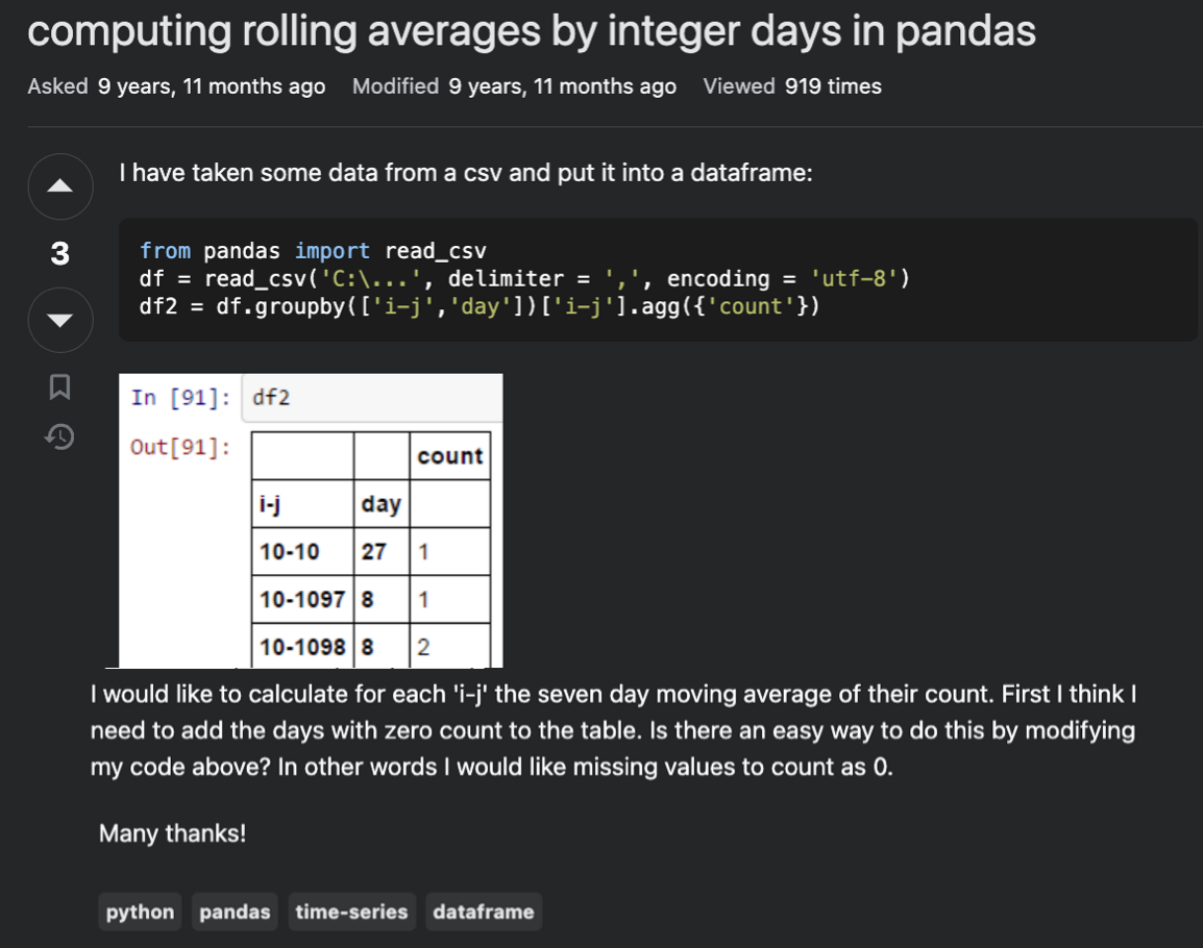}
    \caption{Stack Overflow post tagged with the Python programming language and containing a code snippet with a library import.}
    \label{fig:example-image}
\end{figure}

Our regular expressions handle multiple ways that users can import libraries in the different programming languages. For example, in the Python programming language there are multiple valid ways to import libraries:

\begin{lstlisting}[language=Python], caption={Multiple valid ways of importing libraries in the Python programming language.}
import pandas as pd
from numpy import array
import glob
import requests, math
\end{lstlisting}

\subsection*{Checking validity}
Stack Overflow code snippets are generally not valid programs, and there is no guarantee that import statements we detect refer to actual libraries. For example, a post may refer to libraries ``foo'', ``bar'', and ``foobar'' as part of an illustrative example. Our subsequent analyses rely on a relatively accurate identification of real imports of libraries. Thus, we checked how many library imports we observed in Python posts correspond to actual libraries recorded in the Python Package Index (PyPI) and the list of Python standard library packages (extracted for Python 2.7.18 and Python 3.12.6). We find that 85\% of the Python libraries imported refer to libraries in the canonical list. We assume that the rate of error is similar across other ecosystems.

\subsection*{Quantifying novelty}
We consider a library import a novelty in the respective programming language the first time it appears in a post. We consider two libraries a novel pair the first time they co-occur in a post. Note that, in order to form a novel pair, neither of the imported libraries has to be a novelty. For example, in the hypothetical scenario below, the posts contain 2, 1, and 0 novel libraries, respectively, assuming the posts appear in left-to-right order and no prior posts import any of the mentioned libraries. All of the listed posts introduce a novel pair.

In our analyses of novelty, we filter out libraries that were imported less than 10 times in subsequent posts. Our decision is driven by the goal of avoiding the analysis of small personal projects that did not achieve widespread adoption as innovations. Similarly, we only consider pairs of libraries a novel pair in our analyses if both libraries of the pair were imported at least 10 times throughout our dataset.

\begin{center}
\begin{figure}
\begin{minipage}[t]{0.15\textwidth}
    \centering
   \begin{lstlisting}[language=Python]
import os
import sys
\end{lstlisting}
    \captionof*{figure}{Post 1}
\end{minipage}
\hspace{0.1cm}
\begin{minipage}[t]{0.15\textwidth}
    \centering
   \begin{lstlisting}[language=Python]
import random
import sys
\end{lstlisting}
    \captionof*{figure}{Post 2}
\end{minipage}
\hspace{0.1cm}
\begin{minipage}[t]{0.15\textwidth}
    \centering
   \begin{lstlisting}[language=Python]
import os
import random
\end{lstlisting}
    \captionof*{figure}{Post 3}
\end{minipage}
\caption{Toy example of three sequential posts with import statements in the Python programming language. The first post contains two novel imports, the second contains one, and the third post contains none. The first post contains one novel combination of imports (os, sys), the second contains one novel combination (random, sys), and the third also contains one novel combination (os, random).}
\end{figure}
\end{center}

\subsection*{User characteristics}

We consider two characteristics of posting users that may relate to their likelihood of posting novelties: their experience and geographic location. We define a user's experience at the time of a post by the number of posts they had made previously in the programming language of the post. To infer a posting user's country, we geocode the location they provide in their profile pages using the Microsoft Bing Maps API, an accurate tool which handles multiple languages and representations of locations \cite{wachs2022geography}. We infer a country for 38\% of posting users.

\section{Results}

In this section we present our findings. First we show how simple and combinatorial novelties emerge per post in each of the twelve ecosystems. We then report the concentration of library use across posts. Finally, we report on the relationships between posting user characteristics (experience and location) and novelties.

\subsection*{Novelties}
\begin{figure*}
    \centering
    \includegraphics[width=0.86\textwidth]{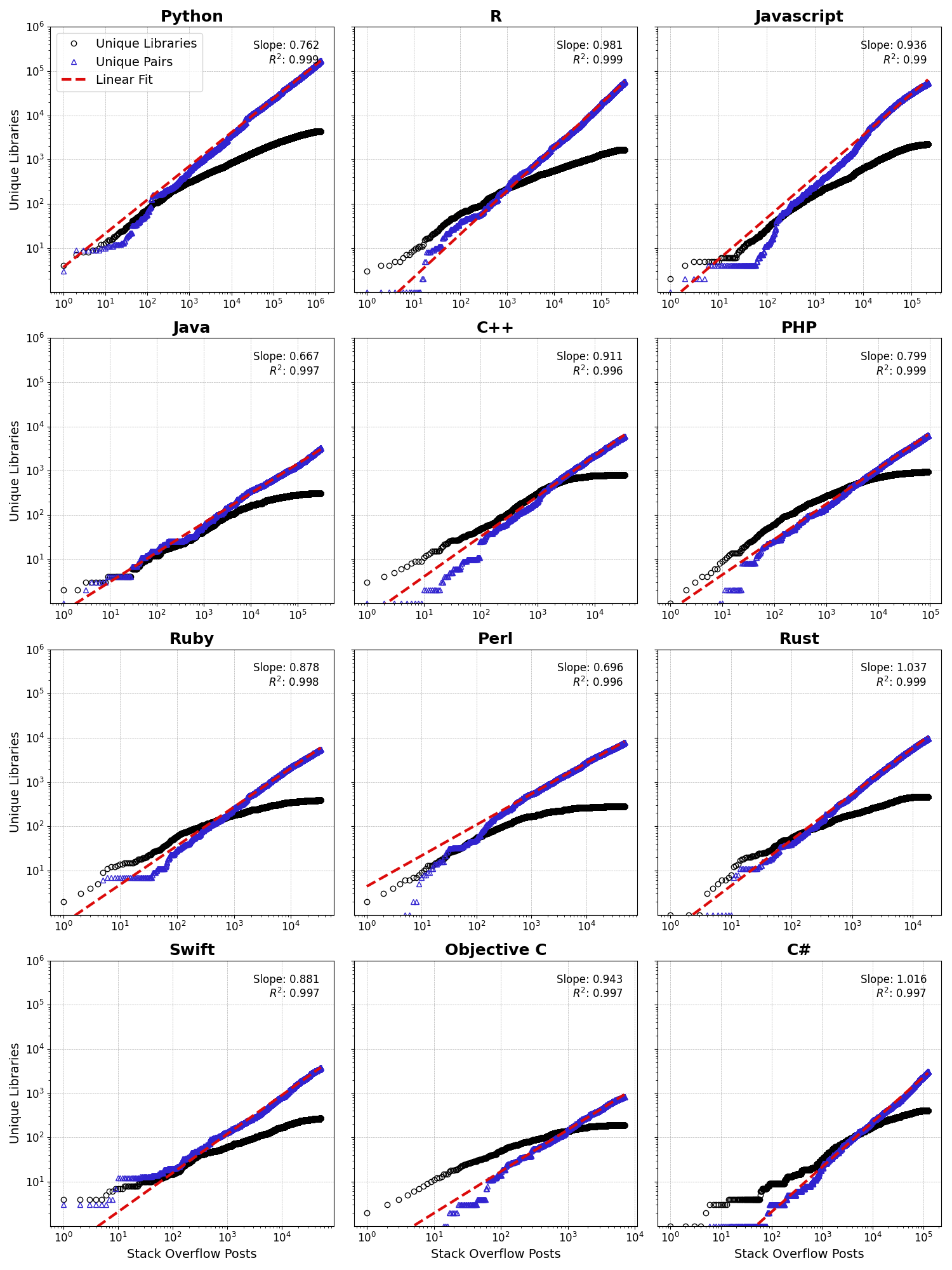}
    \caption{Rates of simple and combinatorial novelties of library imports in Stack Overflow posts across programming languages.}
    \label{fig:rate_nov}
\end{figure*}

In Figure~\ref{fig:rate_nov} we report data on novel library imports and combinations of library imports per post. Novelties in all of the studied programming languages followed a sub-linear growth pattern. In other words, as these systems grow, fewer new libraries are introduced for the same level of activity. At the same time, the rate of combinatorial novelties grows at a remarkably steady rate in all ecosystems: linear fits are highly accurate. These two patterns replicate earlier findings on the introduction of new patent classes in a dataset of over 200 years of patents from the US \cite{youn2015invention}. There too, new classes emerged at an ever-slower rate, while novel pairs of classes emerged at a constant rate per patent across the entire dataset. 

This finding has a few interesting implications. The first is that innovation in mature OSS ecosystems happens by combining libraries. Second, within an ecosystem, the rate at which users explore new combinations of libraries or exploit existing combinations is roughly constant over time. The third implication, which we explore in greater detail in the subsequent subsection, is that, relative to the number of posts made, the number of libraries in an ecosystem is relatively small. In other words, we expect use of libraries to be significantly concentrated in a few key libraries.

\subsection*{Concentration of use}
\begin{figure}
    \centering
    \includegraphics[width=0.5\textwidth]{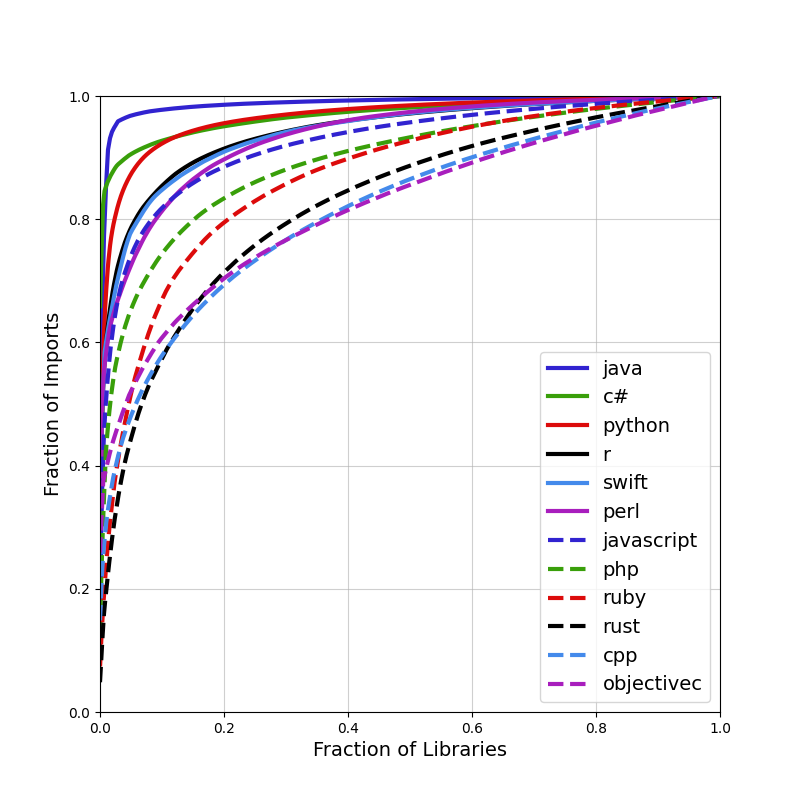}
    \caption{The fraction of total imports in specific programming languages made of the fraction of the most commonly imported libraries. For example, imports of the 7\% most frequently imported Python libraries account for 90\% of all imports.}
    \label{fig:concentration}
\end{figure}

Given that the rate of new libraries introduced per post universally declines across all programming languages we study, we suspect that the use of libraries across all posts is highly concentrated \cite{tria2018zipf}. To examine this hypothesized concentration, we draw Pareto curves \cite{newman2005power} in Figure~\ref{fig:concentration}. These curves are often used to visualize the distribution of wealth in an economy, and are related to the ``80-20'' rule.

In our context, they report the cumulative share of all imports made in an ecosystem by individual libraries, recorded in descending order by number of imports. If each library were imported an equal number of times, we would expect a diagonal ($y=x$) line. If all imports were concentrated in a single library, we would observe a vertical line from (0,0) to (0,1), then a horizontal line from (0,1) to (1,1). Generally, the closer the observed curve is to the top left corner of the plot, the more concentrated the imports are among a few libraries.

Figure~\ref{fig:concentration} confirms that imports are intensely concentrated among the top libraries in most ecosystems. For example, 90\% of all imports in Python are of the 7\% most frequently imported libraries. To facilitate interpretation, we report several similar statistics for all language ecosystems in Table~\ref{tab:pareto_table}.

\begin{table}
\centering
\begin{tabular}{|l|r|r|r|r|r|}
\hline
\textbf{Language} & \textbf{Libraries} & \textbf{Imports} & \textbf{50\%} & \textbf{80\%} & \textbf{90\%} \\
\hline
Python     & 4,366  & 2,469,502 & 0.4  & 2.4  & 7.0  \\
R          & 1,669  & 528,278   & 0.4  & 5.8  & 16.6 \\
Objective-C & 192    & 8,767     & 4.2  & 37.0 & 62.5 \\
Javascript & 2,205  & 413,073   & 1.0  & 8.4  & 23.8 \\
Java       & 310    & 508,724   & 0.0  & 1.0  & 1.3  \\
C++       & 801    & 44,341    & 5.9  & 36.0 & 59.9 \\
PHP       & 936    & 106,829   & 1.7  & 15.3 & 36.1 \\
Ruby       & 385    & 48,770    & 4.7  & 20.8 & 40.8 \\
Perl       & 279    & 119,065   & 0.4  & 9.3  & 20.8 \\
Rust       & 465    & 30,995    & 6.9  & 31.2 & 54.0 \\
Swift      & 268    & 83,549    & 0.7  & 6.3  & 18.3 \\
C\#        & 409    & 169,781   & 0.0  & 0.5  & 4.4  \\
\hline
\end{tabular}
\caption{Library import frequencies per programming language ecosystem. We report the unique number of libraries, frequency of their imports, and the share of top libraries responsible for 50\%, 80\% and 90\% of imports. For example, the top 0.4\% of libraries account for 50\% of all imports in Python posts to Stack Overflow.
}
\label{tab:pareto_table}
\end{table}

These results suggest that a few libraries are used to solve most problems in any given ecosystem. This has consequences for maintenance of these ecosystems. On the one hand, attention and resources can effectively be concentrated on the most used libraries. On the other hand, errors and defects in these key libraries can have a very broad impact on end users. This analysis of library use complements previous work on the distribution of technical dependencies between libraries, which similarly suggests that a few key libraries are highly depended on \cite{decan2019empirical}.

\subsection*{Users}
In our final set of analyses, we focus on user attributes which correlate with innovation. We first investigate how prior user experience in a language ecosystem correlates with the likelihood that a user will post a novel library or pair of libraries. On the one hand, new users may bring new perspectives, problems, and solutions to an ecosystem. On the other hand, experienced users may be the most knowledgeable about libraries available in an ecosystem, and may be dealing with more complex problems that require highly specialized libraries or combinations. If the former effect dominates, we would expect new users to be more likely to import novel libraries. If the latter is true, we would expect to see novelties appear more frequently on posts made by expert users.

\begin{figure*}
    \centering
    \includegraphics[width=0.8\textwidth]{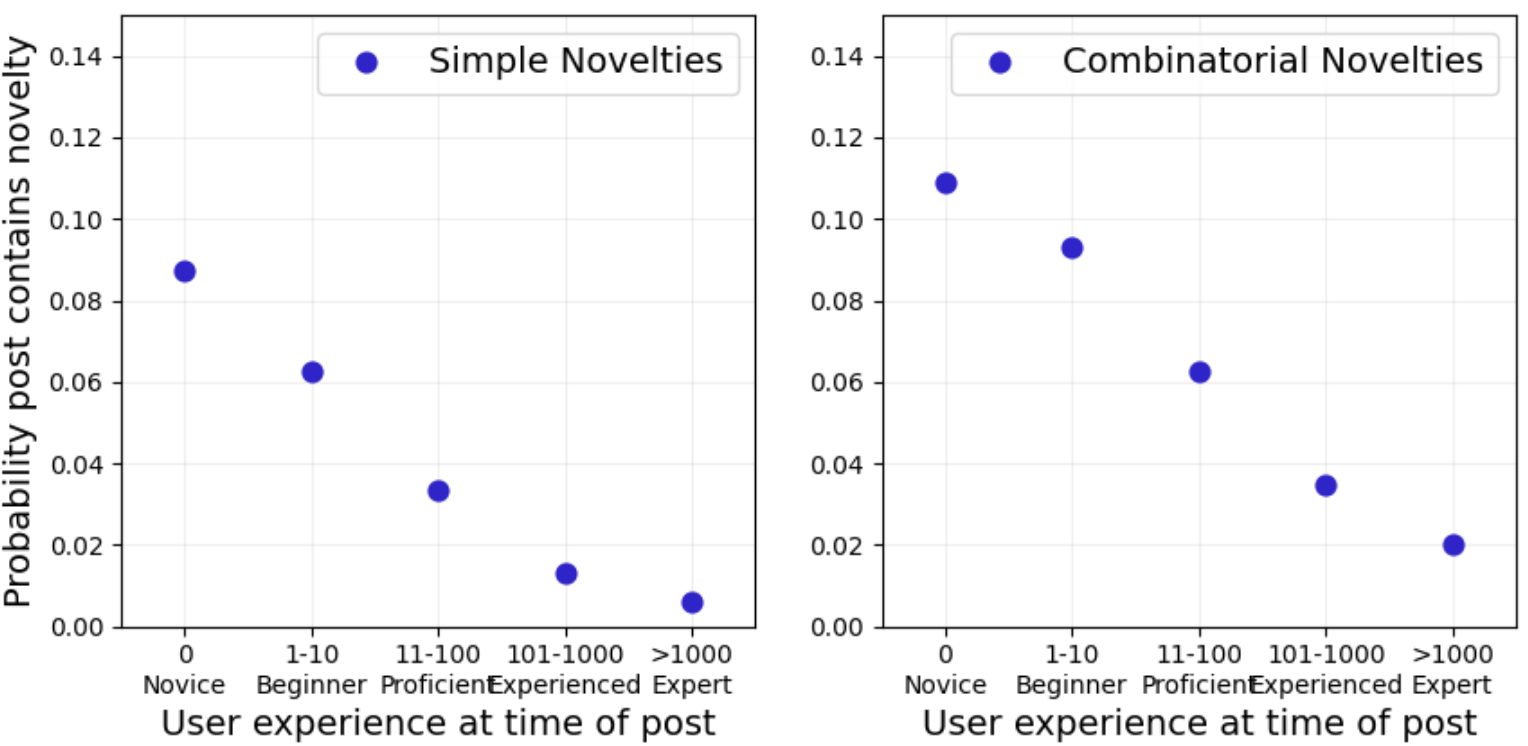}
    \caption{The likelihood that a post contains a novel library or pair of libraries, respectively, as a function of the user's previous posting experience at the time of the post. Note that we define user experience and novelty of posts at the programming language level, reporting pooled estimates here.}
    \label{fig:user_exp_total}
\end{figure*}
To evaluate these competing predictions, for each post we quantify user experience as the prior number of posts made by that user in that ecosystem. We log-bin users into five categories of experience and plot the likelihood that a post by that user contains a novel library or pair in Figure~\ref{fig:user_exp_total}. This plot shows estimates pooled across all language ecosystem; we find similar results in all individual ecosystems, reported in the Appendix.

For both kinds of novelty we observe a striking pattern: new users are significantly more likely to import novel libraries and combinations of novel libraries. Across all languages pooled together, a beginner (having made 1-10 posts previously) is about four times more likely to make a post with a new library than an experienced user (101-1000 prior posts), and three times more likely to make a post with a combinatorial novelty. In other words, new users are introducing new ideas to OSS ecosystems. This finding also has implications for ecosystem sustainability, suggesting that the onboarding and inclusion of new contributors and users pays off in terms of innovation outcomes.

 We check the robustness of this finding in a variety of ways. We first dropped novel libraries and pairs of libraries if those libraries or pairs were not used frequently in subsequent posts. Indeed new users may be more likely to import unsophisticated, less useful, or incoherent libraries or combinations of libraries. However, we found no substantive difference when keeping only novelties that were more broadly adopted (used in posts at least 1,000 times). We also found a similar pattern focusing only on posts made in specific years, alleviating the concern that our results are an artifact of the concentration of beginner status users and novelties in posts at the beginning of our dataset.

\begin{figure*}
    \centering
    \includegraphics[width=\textwidth]{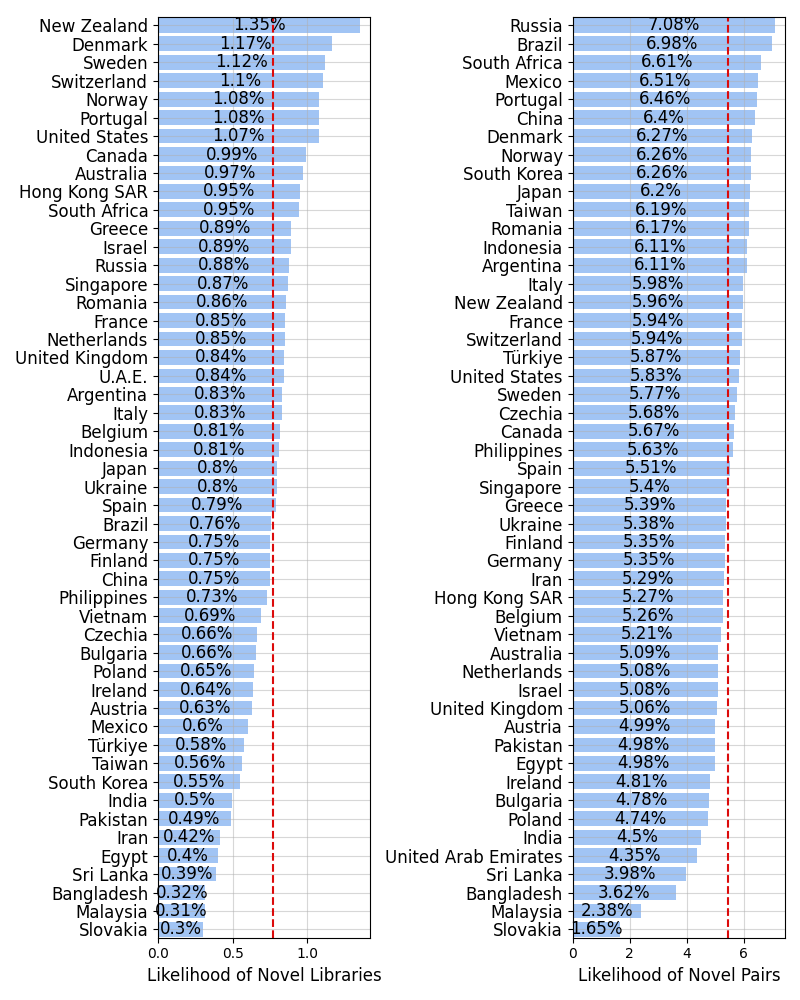}
    \caption{Rates of novelty and combinatorial novelty in posts by users geolocated to different countries. Red vertical lines indicate global averages.}
    \label{fig:geo}
\end{figure*}

Finally, the second source of user-level heterogeneity in innovation we investigated was geography. In Figure \ref{fig:geo} we report the likelihood that posts made by users from different countries contain novelties. Note that we report countries which we can associate to at least 1000 posts. We observe heterogeneities between countries: the rate of both simple and combinatorial novelty is several times greater in the leading countries versus the lagging countries. At the same time, we observe significant geographic diversity in signals of rates of innovation: no small, coherent group of countries dominates the others. This suggests that despite the previously documented spatial concentration of OSS activity \cite{wachs2022geography}, new ideas can come from anywhere.

\section{Discussion}

Our study investigates the dynamics of innovation within OSS ecosystems, revealing several key findings. We observed that the introduction of new libraries follows a sub-linear growth pattern, indicating a slowdown in creating entirely new libraries as ecosystems mature. Conversely, combinatorial innovation—where existing libraries are reused and combined in novel ways—exhibits steady linear growth. This trend suggests that mature OSS ecosystems tend to innovate by combining existing components rather than developing new foundational ones. Our analysis highlights a significant concentration in library usage, with a few key libraries becoming highly integrated across various projects. We found that innovation was more likely to be found in posts by newer users. Finally, the geography of innovative users was found to be quite diverse. 

Here we discuss implications of our findings, focus on two aspects: OSS ecosystem sustainability and the dynamics of innovation in OSS compared with other systems. We then elaborate on the limitations of our work and suggest directions for future research.

\paragraph{Implications for OSS Ecosystem Sustainability}

The concentrated use of certain libraries has both positive and negative implications for the sustainability of OSS ecosystems \cite{decan2019empirical, miller2023we}. On the positive side, heavily utilized core libraries can enhance stability, benefiting from robust community support and well-maintained codebases \cite{avelino2019abandonment}. This shared foundation fosters consistent practices among developers. However, the reliance on a few key libraries also poses risks. If a widely used library becomes deprecated or encounters significant issues, it could threaten the entire ecosystem \cite{zimmermann2019small}. Another challenge our findings present is that combinations of libraries are used at a steadily growing rate in all ecosystems. Hence, ensuring that combinations of libraries, which may not depend on each other in a formal sense, work together is an especially daunting task in light of this growth pattern.

At the user level, we showed that new contributors play a pivotal role in driving innovation within OSS ecosystems \cite{trinkenreich2020hidden, gerosa2021shifting}. They often introduce novel combinations of libraries and experiment with approaches that may not occur to seasoned contributors \cite{fang2024novelty}. This underscores the importance of community practices that lower barriers to entry and encourage diverse participation \cite{steinmacher2015social, mendez2018open}. Improving documentation, streamlining onboarding processes, and offering mentorship can help harness the innovative potential of newcomers \cite{trinkenreich2020hidden}. Creating an inclusive and supportive environment where new ideas are valued is crucial for sustaining their contributions over time \cite{gerosa2021shifting, mendez2018open}.

Our findings also reveal substantial geographical diversity among OSS library contributors. Signals of new library and library-combination use come from all over the world, indicating that innovation in use is not confined to traditional software hubs \cite{musseau2022open}. This diversity brings a range of perspectives and problem-solving approaches, enhancing the resilience and adaptability of OSS ecosystems. However, our findings refer to the rates of innovation as a function of activity, and geographic disparities in activity certainly exist \cite{steinmacher2015social}. This suggests that increasing participation in and use of OSS can spur innovation in the overall system. Addressing this gap requires targeted outreach and support initiatives to empower contributors from emerging markets and underrepresented areas \cite{mendez2018open, trinkenreich2020hidden}. 

In sum, our work suggests a few practical ideas for improving OSS ecosystem sustainability. First, prioritizing support for widely used libraries is crucial to ensure they remain well-maintained and resilient; this can be achieved through funding and community-driven initiatives \cite{avelino2019abandonment, eghbal2016roads}; moreover, maintenance should also be understood as supporting the combinations of libraries. Second, fostering the involvement of new contributors is essential for ongoing innovation \cite{trinkenreich2020hidden, gerosa2021shifting}. Implementing structured onboarding programs, providing mentorship, and offering resources to guide newcomers can facilitate their integration and contribution \cite{steinmacher2015social, trinkenreich2020hidden}.

\paragraph*{Innovation in OSS and other systems}

We have shown that the patterns of innovation observed in OSS ecosystems are similar to those in other fields like technological patents and scientific research, where significant advances often result from novel recombinations of existing elements rather than entirely new creations \cite{weitzman1998recombinant,youn2015invention}. This parallel suggests that insights from evolutionary theories of the economy may be applied to understand OSS ecosystems \cite{nelson1985evolutionary}. In that tradition, firms adapt and imitate each other, driven by processes of variation, selection, and retention that influence their competitive advantage and sustainability. Similarly, OSS ecosystems evolve through cycles of new library introductions, novel code combinations, and the community's selection of effective tools and practices. While software engineering literature has explored evolutionary concepts and the development of software ecosystems, it often emphasizes technical growth, change propagation, and dependency management \cite{mens2014evolving} or within-library evolution \cite{kemerer1999empirical}. 

For instance, in the context of patents we know that social and collaboration network positions of individuals influence their ability to combine ingredients in a productive way \cite{kogut1992knowledge}, and that more complex and unorthodox combinations lead to higher variations in outcomes \cite{fleming2001recombinant}. Similar studies of academic research papers suggests that actually realized combinations of ideas tend to be repetitive and imitative, and that the most successful papers tend to mix in just a few surprising ingredients \cite{uzzi2013atypical}. Indeed recent work in software engineering references this tradition showing combinatorial innovation of libraries within software projects predicts success outcomes \cite{fang2024novelty}. Likewise, collaboration network topology within software ecosystems, like in patenting, play a key role in the spread and adoption of new ingredients (libraries) \cite{fang2024weaktiesexplainopen}.

\subsection*{Limitations and Future Work}

While our study provides insights into the dynamics of innovation within OSS ecosystems, it is important to acknowledge its limitations, particularly regarding our reliance on Stack Overflow as the primary data source \cite{baltes2019sotorrent}. Stack Overflow offers a rich repository of developer interactions, which we utilize as signals to infer rates of innovation and the underlying dynamics of library usage \cite{wu2019developers}. However, we do not consider the specific posts or discussions on Stack Overflow as direct instances of innovation themselves. Instead, they serve as proxies indicating how developers engage with and combine existing libraries.

This reliance on a single platform introduces potential biases. Stack Overflow tends to highlight issues and topics that generate widespread interest or challenges among developers, possibly overlooking smaller-scale innovations or niche libraries that do not elicit as much public discussion \cite{wu2019developers}. Consequently, our analysis may underrepresent innovative activities occurring in less-discussed areas of the OSS ecosystem. Additionally, not all developers use Stack Overflow uniformly; some may prefer other forums, private communications, or may not seek external help at all, leading to gaps in the data captured.

To address these limitations, future research should incorporate a more diverse set of data sources. Direct analysis of code repositories, commit histories, package managers, and issue trackers could provide a more comprehensive view of library usage and innovation patterns. These sources can capture actual coding activities and library integrations that may not be discussed publicly. Combining multiple data sources would help mitigate the biases inherent in relying solely on Stack Overflow.

Furthermore, our study focuses on observable signals of innovation and may not fully capture the qualitative aspects of how developers innovate. Understanding the motivations, decision-making processes, and collaborative efforts behind library usage requires qualitative methods such as surveys or interviews. Incorporating these approaches could enrich the analysis and provide deeper insights into the factors driving innovation within OSS ecosystems.

Lastly, the evolving nature of software development practices, including the rise of artificial intelligence and automated code generation tools, may influence innovation dynamics in ways not accounted for in our study \cite{del2024large}. Future investigations should explore how these emerging technologies impact the rates and patterns of innovation, potentially accelerating growth or introducing new forms of collaboration that reshape the OSS ecosystem.

\section{Conclusion}
Software, especially OSS, is often thought of as a highly innovative and creative sector \cite{lakhani2005hackers}. Here we show that the dynamics of innovation in OSS ecosystems are quite similar to the dynamics of innovation in patenting and scientific research. The implications, both for software maintenance and sustainability, and for our general understanding of OSS ecosystem evolution are numerous.  

\section{Data and Code Availability}
The Stack Overflow data dump is available at the Internet Archive: \url{https://archive.org/details/stackexchange}. Processed data (on post-level library use) is available at: \url{https://zenodo.org/uploads/14186439}. Code to replicate our analyses is available here:\url{https://github.com/MeszarosGabor/SO_Post_Analyzer}.

\section{Acknowledgments}
We thank participants of the Danube Workshop on Software and the Digital Economy for helpful comments and suggestions. JW acknowledges support from the Hungarian National Scientific Fund (OTKA FK 145960). GM is funded by the European Union under Horizon EU project LearnData (101086712). The authors acknowledge use of the HUN-REN Cloud \cite{heder2022past} in the ``Geographies of Creation, Learning and Use in Software'' project.

\printbibliography

@inproceedings{baltes2019sotorrent,
  title={Sotorrent: Studying the origin, evolution, and usage of stack overflow code snippets},
  author={Baltes, Sebastian and Treude, Christoph and Diehl, Stephan},
  booktitle={2019 IEEE/ACM 16th International Conference on Mining Software Repositories (MSR)},
  pages={191--194},
  year={2019},
  organization={IEEE}
}

@article{charette2021car,
  title={How Software is Eating the Car},
  author={Charette, Robert N},
  journal={IEEE Spectrum},
  year={2021}
}

@article{youn2015invention,
  title={Invention as a combinatorial process: evidence from US patents},
  author={Youn, Hyejin and Strumsky, Deborah and Bettencourt, Luis MA and Lobo, Jos{\'e}},
  journal={Journal of the Royal Society interface},
  volume={12},
  number={106},
  pages={20150272},
  year={2015},
  publisher={The Royal Society}
}

@article{stojkoski2024estimating,
  title={Estimating digital product trade through corporate revenue data},
  author={Stojkoski, Viktor and Koch, Philipp and Coll, Eva and Hidalgo, C{\'e}sar A},
  journal={Nature Communications},
  volume={15},
  number={1},
  pages={5262},
  year={2024},
  publisher={Nature Publishing Group UK London}
}

@inproceedings{dabbish2012social,
  title={Social coding in GitHub: transparency and collaboration in an open software repository},
  author={Dabbish, Laura and Stuart, Colleen and Tsay, Jason and Herbsleb, Jim},
  booktitle={Proceedings of the ACM 2012 conference on computer supported cooperative work},
  pages={1277--1286},
  year={2012}
}

@book{eghbal2020working,
  title={Working in public: the making and maintenance of open source software},
  author={Eghbal, Nadia},
  year={2020},
  publisher={Stripe Press}
}

@article{newman2005power,
  title={Power laws, Pareto distributions and Zipf's law},
  author={Newman, Mark EJ},
  journal={Contemporary physics},
  volume={46},
  number={5},
  pages={323--351},
  year={2005},
  publisher={Taylor \& Francis}
}

@article{frakes2005software,
  title={Software reuse research: Status and future},
  author={Frakes, William B and Kang, Kyo},
  journal={IEEE Transactions on Software Engineering},
  volume={31},
  number={7},
  pages={529--536},
  year={2005},
  publisher={Ieee}
}

@article{louridas2008power,
  title={Power laws in software},
  author={Louridas, Panagiotis and Spinellis, Diomidis and Vlachos, Vasileios},
  journal={ACM Transactions on Software Engineering and Methodology (TOSEM)},
  volume={18},
  number={1},
  pages={1--26},
  year={2008},
  publisher={ACM New York, NY, USA}
}

@article{dietrich2007optimal,
  title={The optimal age to start a revolution},
  author={Dietrich, Arne and Srinivasan, Narayanan},
  journal={The Journal of Creative Behavior},
  volume={41},
  number={1},
  pages={54--74},
  year={2007},
  publisher={Wiley Online Library}
}

@inproceedings{decan2018impact,
  title={On the impact of security vulnerabilities in the npm package dependency network},
  author={Decan, Alexandre and Mens, Tom and Constantinou, Eleni},
  booktitle={Proceedings of the 15th international conference on mining software repositories},
  pages={181--191},
  year={2018}
}

@article{valverde2002scale,
  title={Scale-free networks from optimal design},
  author={Valverde, Sergi and Cancho, R Ferrer and Sole, Richard V},
  journal={Europhysics Letters},
  volume={60},
  number={4},
  pages={512},
  year={2002},
  publisher={IOP Publishing}
}

@article{tria2018zipf,
  title={Zipf’s, Heaps’ and Taylor’s Laws are Determined by the Expansion into the Adjacent Possible},
  author={Tria, Francesca and Loreto, Vittorio and Servedio, Vito DP},
  journal={Entropy},
  volume={20},
  number={10},
  pages={752},
  year={2018},
  publisher={MDPI}
}

@article{zhu2013geography,
  title={Geography and similarity of regional cuisines in China},
  author={Zhu, Yu-Xiao and Huang, Junming and Zhang, Zi-Ke and Zhang, Qian-Ming and Zhou, Tao and Ahn, Yong-Yeol},
  journal={PloS one},
  volume={8},
  number={11},
  pages={e79161},
  year={2013},
  publisher={Public Library of Science San Francisco, USA}
}

@article{tria2014dynamics,
  title={The dynamics of correlated novelties},
  author={Tria, Francesca and Loreto, Vittorio and Servedio, Vito Domenico Pietro and Strogatz, Steven H},
  journal={Scientific Reports},
  volume={4},
  number={1},
  pages={5890},
  year={2014},
  publisher={Nature Publishing Group UK London}
}

@article{cattuto2009collective,
  title={Collective dynamics of social annotation},
  author={Cattuto, Ciro and Barrat, Alain and Baldassarri, Andrea and Schehr, Gregory and Loreto, Vittorio},
  journal={Proceedings of the National Academy of Sciences},
  volume={106},
  number={26},
  pages={10511--10515},
  year={2009},
  publisher={National Acad Sciences}
}

@article{zhang2009discovering,
  title={Discovering power laws in computer programs},
  author={Zhang, Hongyu},
  journal={Information processing \& management},
  volume={45},
  number={4},
  pages={477--483},
  year={2009},
  publisher={Elsevier}
}

@book{baayen2012word,
  title={Word frequency distributions},
  author={Baayen, R Harald},
  volume={18},
  year={2012},
  publisher={Springer Science \& Business Media}
}

@misc{heaps1978information,
  title={Information retrieval: Computational and theoretical aspects},
  author={Heaps, HS},
  year={1978},
  publisher={Academic Press}
}

@book{hausmann2014atlas,
  title={The atlas of economic complexity: Mapping paths to prosperity},
  author={Hausmann, Ricardo and Hidalgo, C{\'e}sar A and Bustos, Sebasti{\'a}n and Coscia, Michele and Simoes, Alexander},
  year={2014},
  publisher={Mit Press}
}

@inproceedings{vadlamani2020studying,
  title={Studying software developer expertise and contributions in Stack Overflow and GitHub},
  author={Vadlamani, Sri Lakshmi and Baysal, Olga},
  booktitle={2020 IEEE International Conference on Software Maintenance and Evolution (ICSME)},
  pages={312--323},
  year={2020},
  organization={IEEE}
}

@article{wu2019developers,
  title={How do developers utilize source code from stack overflow?},
  author={Wu, Yuhao and Wang, Shaowei and Bezemer, Cor-Paul and Inoue, Katsuro},
  journal={Empirical Software Engineering},
  volume={24},
  pages={637--673},
  year={2019},
  publisher={Springer}
}

@inproceedings{vasilescu2014social,
  title={How social Q\&A sites are changing knowledge sharing in open source software communities},
  author={Vasilescu, Bogdan and Serebrenik, Alexander and Devanbu, Prem and Filkov, Vladimir},
  booktitle={Proceedings of the 17th ACM conference on Computer supported cooperative work \& social computing},
  pages={342--354},
  year={2014}
}

@inproceedings{vasilescu2013stackoverflow,
  title={Stackoverflow and github: Associations between software development and crowdsourced knowledge},
  author={Vasilescu, Bogdan and Filkov, Vladimir and Serebrenik, Alexander},
  booktitle={2013 International conference on social computing},
  pages={188--195},
  year={2013},
  organization={IEEE}
}

@book{nelson1985evolutionary,
  title={An evolutionary theory of economic change},
  author={Nelson, Richard R and Winter, Sidney G.},
  year={1985},
  publisher={Harvard University Press}
}

@techreport{gortmaker2024open,
  title={Open Source Software Policy in Industry Equilibrium},
  author={Gortmaker, Jeff},
  year={2024},
  institution={Working paper}
}

@article{kemerer1999empirical,
  title={An empirical approach to studying software evolution},
  author={Kemerer, Chris F. and Slaughter, Sandra},
  journal={IEEE transactions on software engineering},
  volume={25},
  number={4},
  pages={493--509},
  year={1999},
  publisher={IEEE}
}

@book{mens2014evolving,
  title={Evolving Software Systems},
  author={Mens, Tom and Serebrenik, Alexander and Cleve, Anthony and others},
  volume={190},
  year={2014},
  publisher={Springer}
}

@article{del2024large,
  title={Large language models reduce public knowledge sharing on online Q\&A platforms},
  author={del Rio-Chanona, R Maria and Laurentsyeva, Nadzeya and Wachs, Johannes},
  journal={PNAS Nexus},
  volume={3},
  number={9},
  pages={pgae400},
  year={2024},
  publisher={Oxford University Press US}
}

@inproceedings{anderson2012discovering,
  title={{Discovering value from community activity on focused question answering sites: a case study of Stack Overflow}},
  author={Anderson, Ashton and Huttenlocher, Daniel and Kleinberg, Jon and Leskovec, Jure},
  booktitle={Proceedings of the 18th ACM SIGKDD International Conference on Knowledge Discovery and Data Mining},
  year={2012}
}

@article{schumpeter1939business,
  title={Business cycles: A theoretical, historical and statistical analysis of the capitalist process},
  author={Schumpeter, Joseph Alois},
  year={1939},
  publisher={McGraw Hill Book Company}
}

@article{xu2020makes,
  title={What makes geeks tick? A study of stack overflow careers},
  author={Xu, Lei and Nian, Tingting and Cabral, Luis},
  journal={Management Science},
  volume={66},
  number={2},
  pages={587--604},
  year={2020},
  publisher={Informs}
}

@article{trinkenreich2020hidden,
  title={{Hidden figures: Roles and pathways of successful OSS contributors}},
  author={Trinkenreich, Bianca and Guizani, Mariam and Wiese, Igor and Sarma, Anita and Steinmacher, Igor},
  journal={Proceedings of the ACM on human-computer interaction},
  volume={4},
  number={CSCW2},
  pages={1--22},
  year={2020},
  publisher={ACM New York, NY, USA}
}

@article{heder2022past,
  title={The past, present and future of the ELKH cloud},
  author={H{\'e}der, Mih{\'a}ly and Rig{\'o}, Ern{\H{o}} and Medgyesi, Dorottya and Lovas, R{\'o}bert and Tenczer, Szabolcs and Farkas, Attila and Em{\H{o}}di, M{\'a}rk Benj{\'a}min and Kadlecsik, J{\'o}zsef and Kacsuk, P{\'e}ter},
  journal={INFORM{\'A}CI{\'O}S T{\'A}RSADALOM: T{\'A}RSADALOMTUDOM{\'A}NYI FOLY{\'O}IRAT},
  volume={22},
  number={2},
  pages={128--137},
  year={2022}
}

@article{wachs2022geography,
  title={The geography of open source software: evidence from github},
  author={Wachs, Johannes and Nitecki, Mariusz and Schueller, William and Polleres, Axel},
  journal={Technological Forecasting and Social Change},
  volume={176},
  pages={121478},
  year={2022},
  publisher={Elsevier}
}

@inproceedings{gerosa2021shifting,
  title={The shifting sands of motivation: Revisiting what drives contributors in open source},
  author={Gerosa, Marco and Wiese, Igor and Trinkenreich, Bianca and Link, Georg and Robles, Gregorio and Treude, Christoph and Steinmacher, Igor and Sarma, Anita},
  booktitle={2021 IEEE/ACM 43rd International Conference on Software Engineering (ICSE)},
  pages={1046--1058},
  year={2021},
  organization={IEEE}
}

@inproceedings{steinmacher2015social,
  title={Social barriers faced by newcomers placing their first contribution in open source software projects},
  author={Steinmacher, Igor and Conte, Tayana and Gerosa, Marco Aur{\'e}lio and Redmiles, David},
  booktitle={Proceedings of the 18th ACM conference on Computer supported cooperative work \& social computing},
  pages={1379--1392},
  year={2015}
}

@inproceedings{mendez2018open,
  title={Open source barriers to entry, revisited: A sociotechnical perspective},
  author={Mendez, Christopher and Padala, Hema Susmita and Steine-Hanson, Zoe and Hilderbrand, Claudia and Horvath, Amber and Hill, Charles and Simpson, Logan and Patil, Nupoor and Sarma, Anita and Burnett, Margaret},
  booktitle={Proceedings of the 40th International conference on software engineering},
  pages={1004--1015},
  year={2018}
}

@article{uzzi2013atypical,
  title={Atypical combinations and scientific impact},
  author={Uzzi, Brian and Mukherjee, Satyam and Stringer, Michael and Jones, Ben},
  journal={Science},
  year={2013},
  publisher={American Association for the Advancement of Science}
}

@article{schueller2024modeling,
  title={Modeling interconnected social and technical risks in open source software ecosystems},
  author={Schueller, William and Wachs, Johannes},
  journal={Collective Intelligence},
  volume={3},
  number={1},
  pages={26339137241231912},
  year={2024},
  publisher={SAGE Publications Sage UK: London, England}
}

@inproceedings{zimmermann2019small,
  title={Small world with high risks: A study of security threats in the npm ecosystem},
  author={Zimmermann, Markus and Staicu, Cristian-Alexandru and Tenny, Cam and Pradel, Michael},
  booktitle={28th USENIX Security symposium (USENIX security 19)},
  pages={995--1010},
  year={2019}
}

@inproceedings{decan2017empirical,
  title={An empirical comparison of dependency issues in OSS packaging ecosystems},
  author={Decan, Alexandre and Mens, Tom and Claes, Ma{\"e}lick},
  booktitle={2017 IEEE 24th international conference on software analysis, evolution and reengineering (SANER)},
  pages={2--12},
  year={2017},
  organization={IEEE}
}

@inproceedings{pashchenko2018vulnerable,
  title={Vulnerable open source dependencies: Counting those that matter},
  author={Pashchenko, Ivan and Plate, Henrik and Ponta, Serena Elisa and Sabetta, Antonino and Massacci, Fabio},
  booktitle={Proceedings of the 12th ACM/IEEE international symposium on empirical software engineering and measurement},
  pages={1--10},
  year={2018}
}

@inproceedings{avelino2019abandonment,
  title={On the abandonment and survival of open source projects: An empirical investigation},
  author={Avelino, Guilherme and Constantinou, Eleni and Valente, Marco Tulio and Serebrenik, Alexander},
  booktitle={2019 ACM/IEEE International Symposium on Empirical Software Engineering and Measurement (ESEM)},
  year={2019},
  organization={IEEE}
}

@inproceedings{hejderup2022use,
  title={On the Use of Tests for Software Supply Chain Threats},
  author={Hejderup, Joseph},
  booktitle={Proceedings of the 2022 ACM Workshop on Software Supply Chain Offensive Research and Ecosystem Defenses},
  pages={47--49},
  year={2022}
}

@article{geiger2021labor,
  title={The labor of maintaining and scaling free and open-source software projects},
  author={Geiger, R Stuart and Howard, Dorothy and Irani, Lilly},
  journal={Proceedings of the ACM on human-computer interaction},
  volume={5},
  number={CSCW1},
  pages={1--28},
  year={2021},
  publisher={ACM New York, NY, USA}
}

@article{juhasz2024software,
  title={The Software Complexity of Nations},
  author={Juh{\'a}sz, S{\'a}ndor and Wachs, Johannes and Kaminski, Jermain and Hidalgo, C{\'e}sar A},
  journal={arXiv preprint: 2407.13880},
  year={2024}
}

@article{weitzman1998recombinant,
  title={Recombinant growth},
  author={Weitzman, Martin L},
  journal={The Quarterly Journal of Economics},
  volume={113},
  number={2},
  pages={331--360},
  year={1998},
  publisher={MIT Press}
}

@article{kogut1992knowledge,
  title={Knowledge of the firm, combinative capabilities, and the replication of technology},
  author={Kogut, Bruce and Zander, Udo},
  journal={Organization Science},
  year={1992},
  publisher={INFORMS}
}

@article{fleming2001recombinant,
  title={Recombinant uncertainty in technological search},
  author={Fleming, Lee},
  journal={Management science},
  volume={47},
  number={1},
  pages={117--132},
  year={2001},
  publisher={INFORMS}
}

@article{mens2014studying,
  title={Studying evolving software ecosystems based on ecological models},
  author={Mens, Tom and Claes, Ma{\'a}lick and Grosjean, Philippe and Serebrenik, Alexander},
  journal={Evolving software systems},
  pages={297--326},
  year={2014},
  publisher={Springer}
}

@inproceedings{miller2023we,
  title={“We Feel Like We’re Winging It:” A Study on Navigating Open-Source Dependency Abandonment},
  author={Miller, Courtney and K{\"a}stner, Christian and Vasilescu, Bogdan},
  booktitle={Proceedings of the 31st ACM Joint European Software Engineering Conference and Symposium on the Foundations of Software Engineering},
  pages={1281--1293},
  year={2023}
}

@book{mens2023software,
  title={Software Ecosystems: Tooling and Analytics},
  author={Mens, Tom and De Roover, Coen and Cleve, Anthony},
  year={2023},
  publisher={Springer Nature}
}

@book{eghbal2016roads,
  title={Roads and bridges: The unseen labor behind our digital infrastructure},
  author={Eghbal, Nadia},
  year={2016},
  publisher={Ford Foundation}
}

@article{castaldi2018trademark,
  title={To trademark or not to trademark: The case of the creative and cultural industries},
  author={Castaldi, Carolina},
  journal={Research Policy},
  volume={47},
  number={3},
  pages={606--616},
  year={2018},
  publisher={Elsevier}
}

@article{mansfield1986patents,
  title={Patents and innovation: an empirical study},
  author={Mansfield, Edwin},
  journal={Management science},
  volume={32},
  number={2},
  pages={173--181},
  year={1986},
  publisher={INFORMS}
}

@inproceedings{bin2020impact,
  title={The impact of a proposal for innovation measurement in the software industry},
  author={bin Ali, Nauman and Edison, Henry and Torkar, Richard},
  booktitle={Proceedings of the 14th ACM/IEEE International Symposium on Empirical Software Engineering and Measurement (ESEM)},
  pages={1--6},
  year={2020}
}

@inproceedings{musseau2022open,
  title={Is open source eating the world's software? measuring the proportion of open source in proprietary software using Java binaries},
  author={Musseau, Julius and Meyers, John Speed and Sieniawski, George P and Thompson, C Albert and German, Daniel},
  booktitle={Proceedings of the 19th International Conference on Mining Software Repositories},
  pages={561--565},
  year={2022}
}

@article{andreessen2011software,
  title={Why Software is Eating the World},
  author={Andreessen, Marc},
  journal={The Wall Street Journal},
  volume={8},
  pages={20},
  year={2011}
}

@article{chattergoon2022winner,
  title={Winner takes all? Tech clusters, population centers, and the spatial transformation of US invention},
  author={Chattergoon, Brad and Kerr, William R},
  journal={Research Policy},
  volume={51},
  number={2},
  pages={104418},
  year={2022},
  publisher={Elsevier}
}

@article{basili1996reuse,
  title={How reuse influences productivity in object-oriented systems},
  author={Basili, Victor R and Briand, Lionel C and Melo, Walc{\'e}lio L},
  journal={Communications of the ACM},
  volume={39},
  number={10},
  pages={104--116},
  year={1996},
  publisher={ACM New York, NY, USA}
}

@inproceedings{fang2024novelty,
  title={Novelty Begets Popularity, But Curbs Participation-A Macroscopic View of the Python Open-Source Ecosystem},
  author={Fang, Hongbo and Herbsleb, James and Vasilescu, Bogdan},
  booktitle={Proceedings of the 46th IEEE/ACM International Conference on Software Engineering},
  pages={1--11},
  year={2024}
}

@article{korkmaz2024github,
  title={From GitHub to GDP: A framework for measuring open source software innovation},
  author={Korkmaz, Gizem and Calder{\'o}n, J Bayo{\'a}n Santiago and Kramer, Brandon L and Guci, Ledia and Robbins, Carol A},
  journal={Research Policy},
  volume={53},
  number={3},
  pages={104954},
  year={2024},
  publisher={Elsevier}
}

@article{hoffmann2024value,
  title={The Value of Open Source Software},
  author={Hoffmann, Manuel and Nagle, Frank and Zhou, Yanuo},
  journal={Harvard Business School Strategy Unit Working Paper},
  number={24-038},
  year={2024}
}

@misc{fang2024weaktiesexplainopen,
      title={Weak Ties Explain Open Source Innovation}, 
      author={Hongbo Fang and Patrick Park and James Evans and James Herbsleb and Bogdan Vasilescu},
      year={2024},
      eprint={2411.05646},
      archivePrefix={arXiv},
      primaryClass={cs.SE},
      url={https://arxiv.org/abs/2411.05646}, 
}

@book{lakhani2005hackers,
  title={Why hackers do what they do: Understanding motivation and effort in free/open source software projects},
  author={Lakhani, Karim R and Wolf, Robert G},
  year={2005},
  publisher={MIT Press}
}

@article{decan2019empirical,
  title={An empirical comparison of dependency network evolution in seven software packaging ecosystems},
  author={Decan, Alexandre and Mens, Tom and Grosjean, Philippe},
  journal={Empirical Software Engineering},
  volume={24},
  number={1},
  pages={381--416},
  year={2019},
  publisher={Springer}
}

\onecolumn
\section*{Appendix}

\begin{minipage}{\textwidth}
    \centering
    \includegraphics[width=0.8\textwidth]{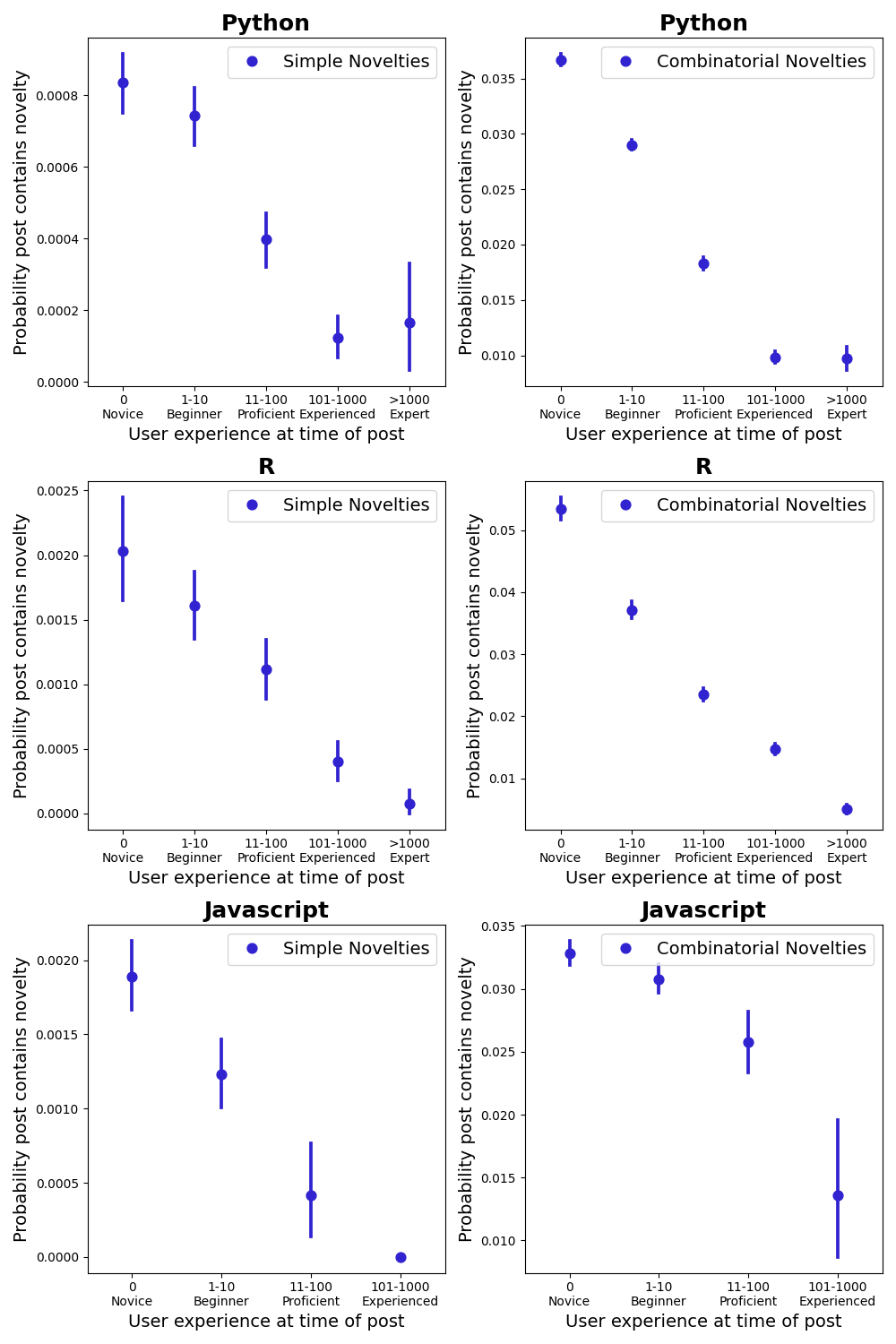}
     \captionof{figure}{User Experience Distribution for Simple and Combinatorial Novelties per Programming Languages I.}
    \label{fig:user_exp_libs_12_a}
\end{minipage}

\begin{figure*}
    \centering
    \includegraphics[width=0.8\textwidth]{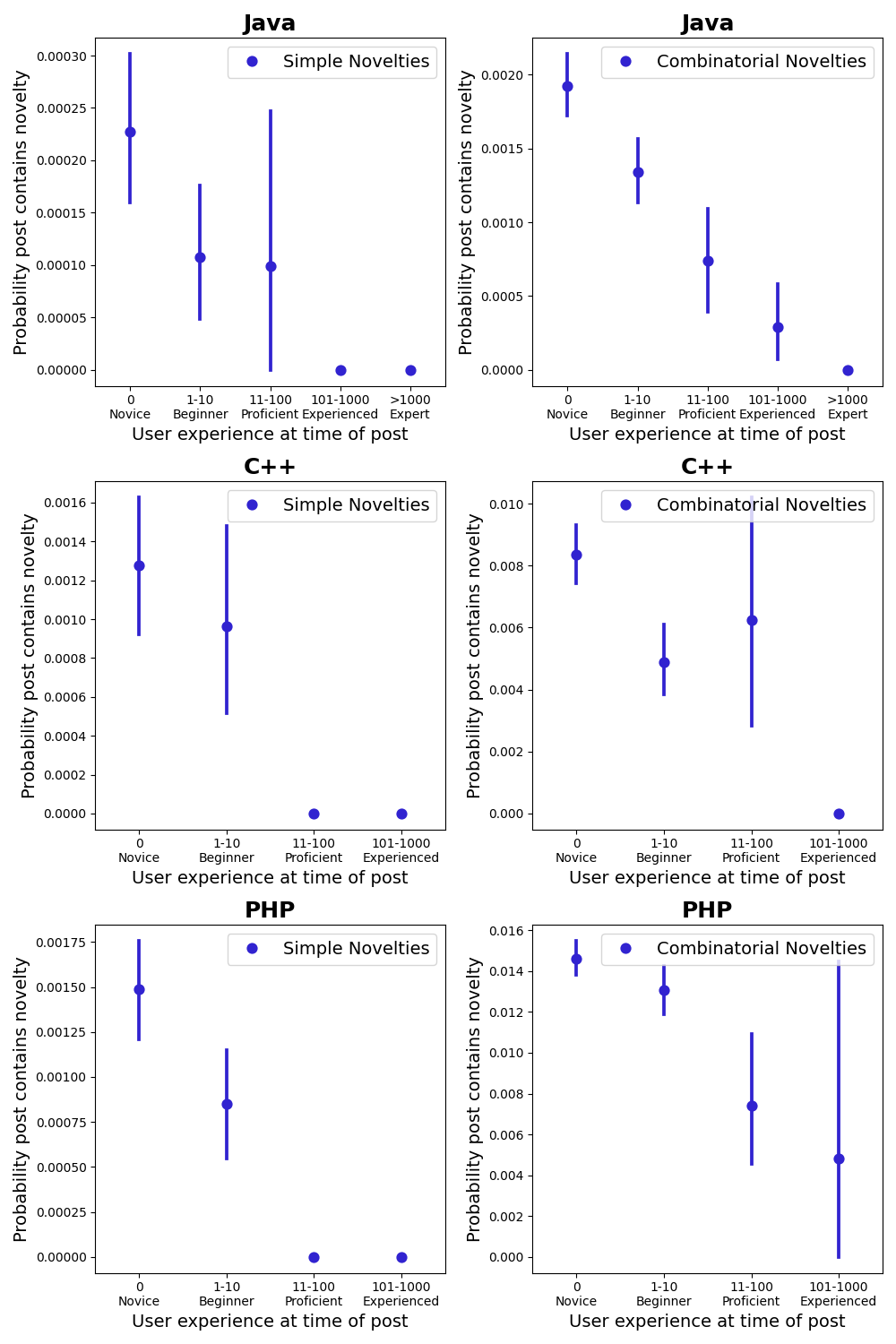}
    \caption{User Experience Distribution for Simple and Combinatorial Novelties per Programming Languages II.}
    \label{fig:user_exp_pairs_12_b}
\end{figure*}
\begin{figure*}
    \centering
    \includegraphics[width=0.8\textwidth]{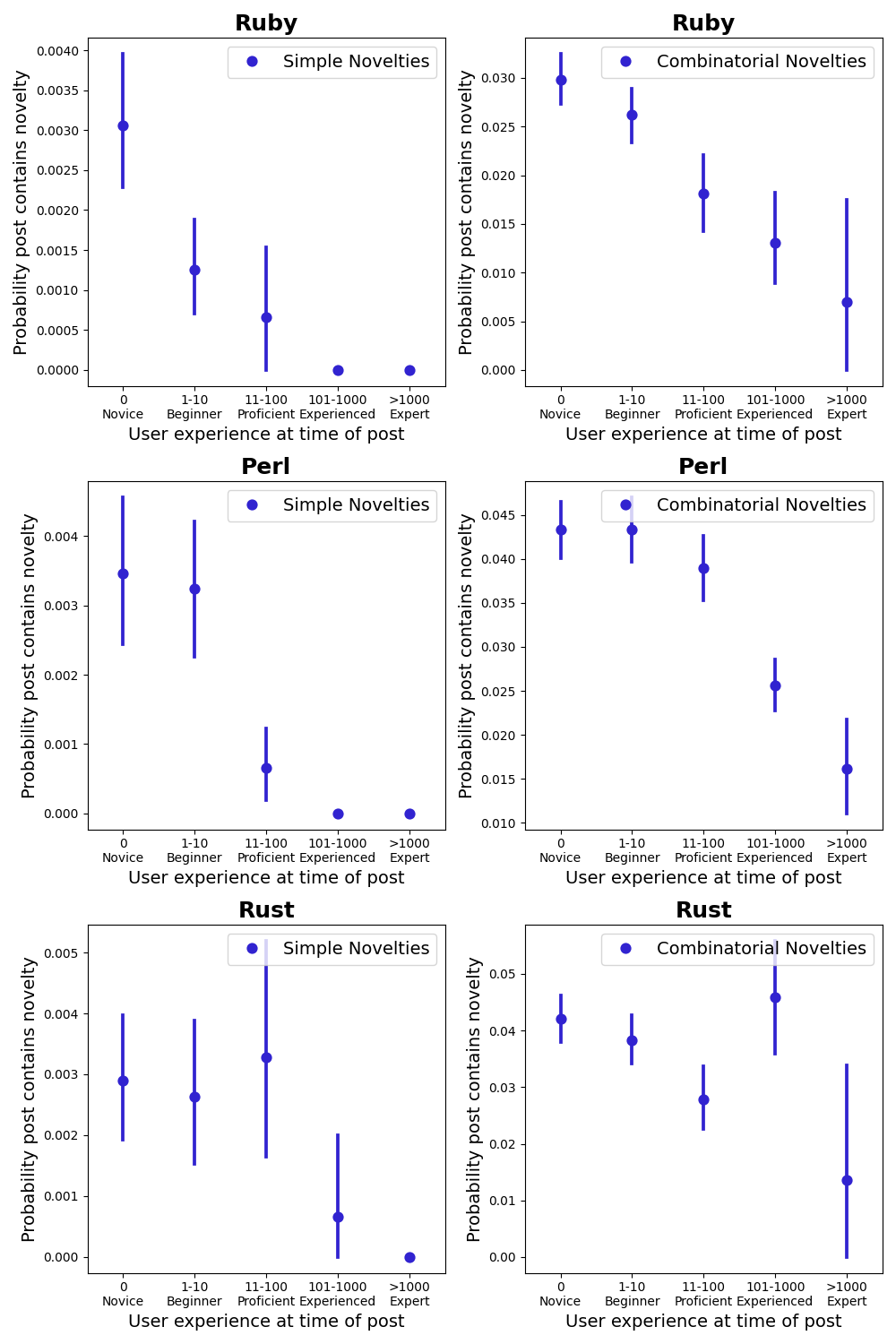}
    \caption{User Experience Distribution for Simple and Combinatorial Novelties per Programming Languages III.}
    \label{fig:user_exp_pairs_12_c}
\end{figure*}
\begin{figure*}
    \centering
    \includegraphics[width=0.8\textwidth]{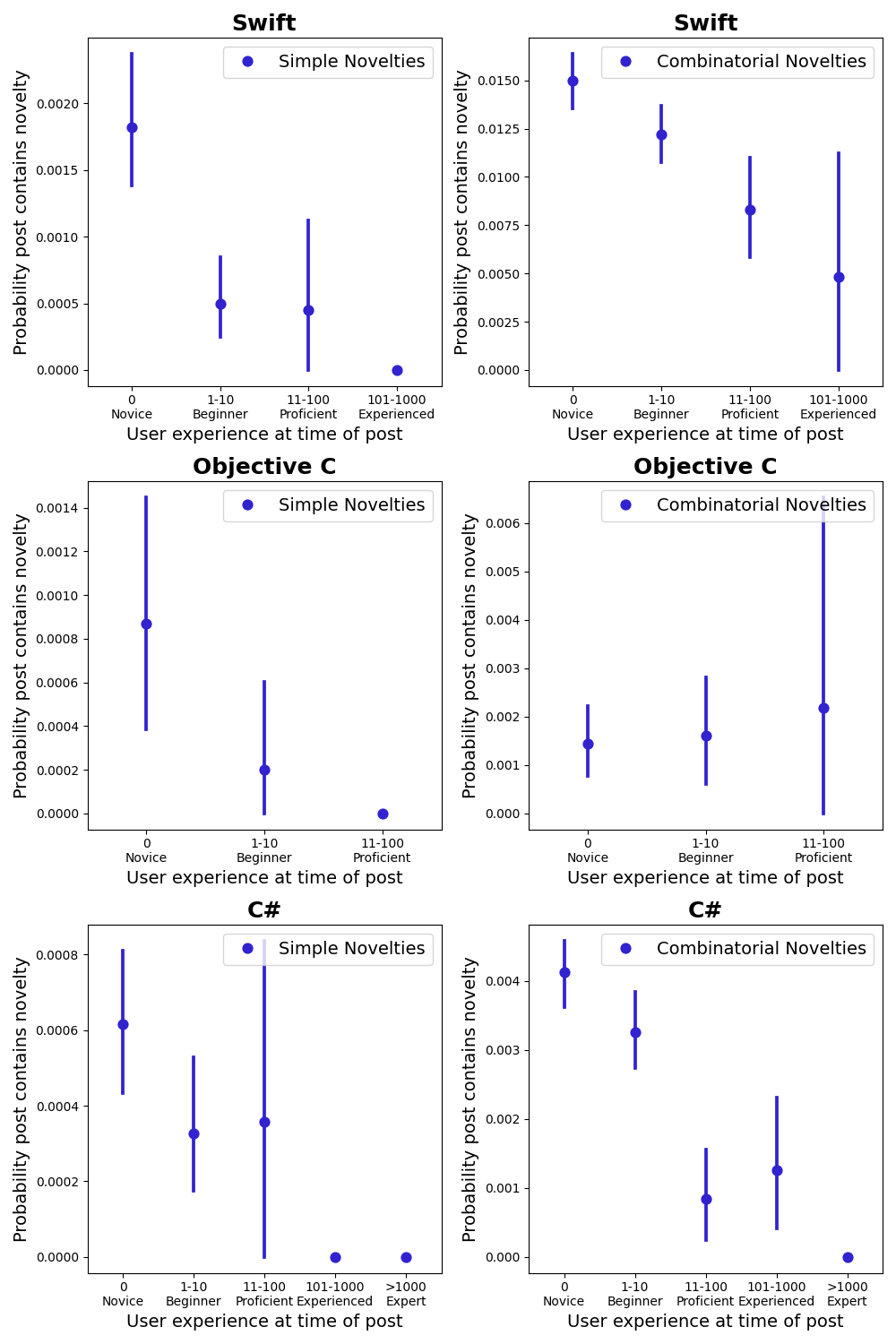}
    \caption{User Experience Distribution for Simple and Combinatorial Novelties per Programming Languages IV.}
    \label{fig:example-user_exp_pairs_12_d}
\end{figure*}
\begin{figure*}
    \centering
    \includegraphics[width=0.8\textwidth]{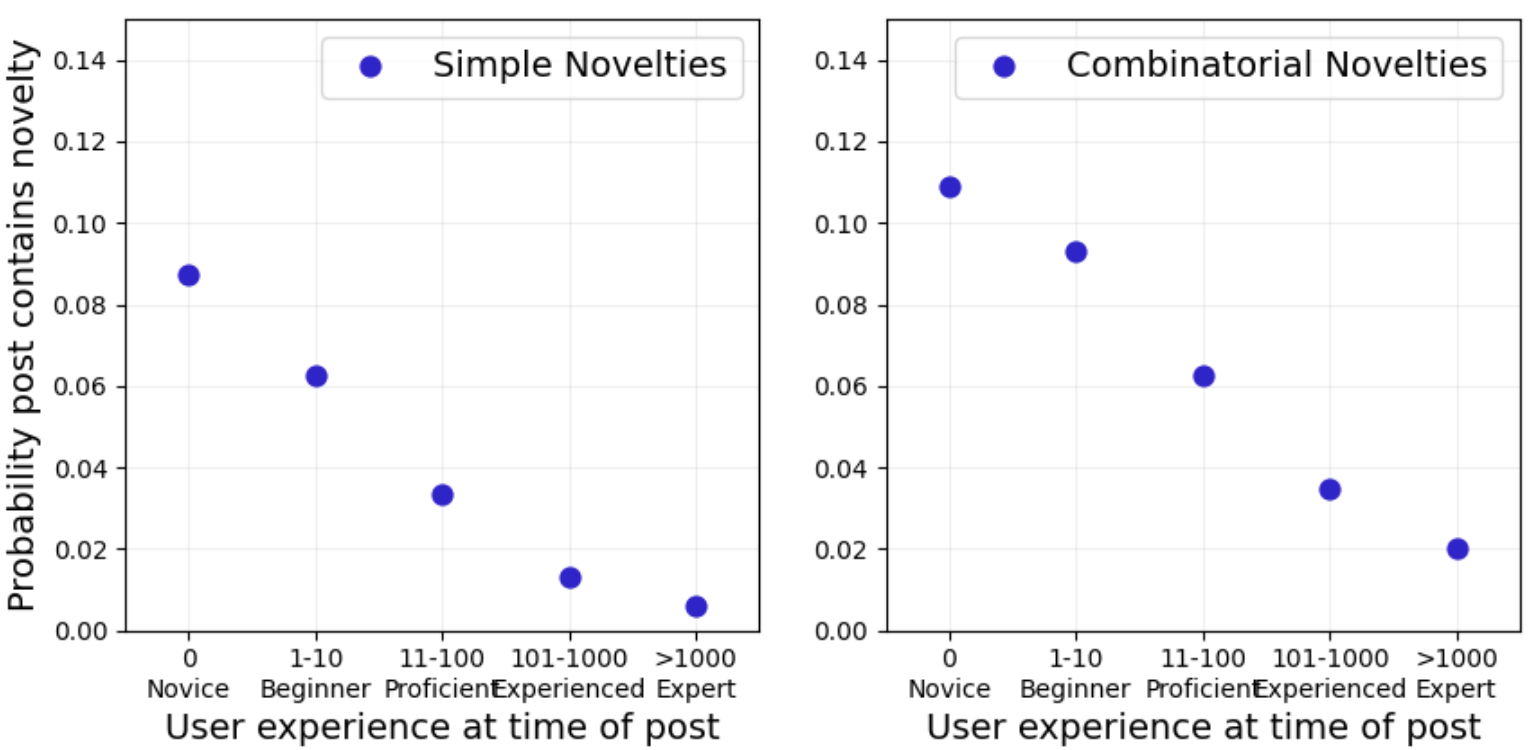}
    \caption{User Experience Distribution for Simple and Combinatorial Novelties among libraries appearing at least 1000 times.}
    \label{fig:example-user_exp_pairs_limit_1000}
\end{figure*}
\begin{figure*}
    \centering
    \includegraphics[width=0.8\textwidth]{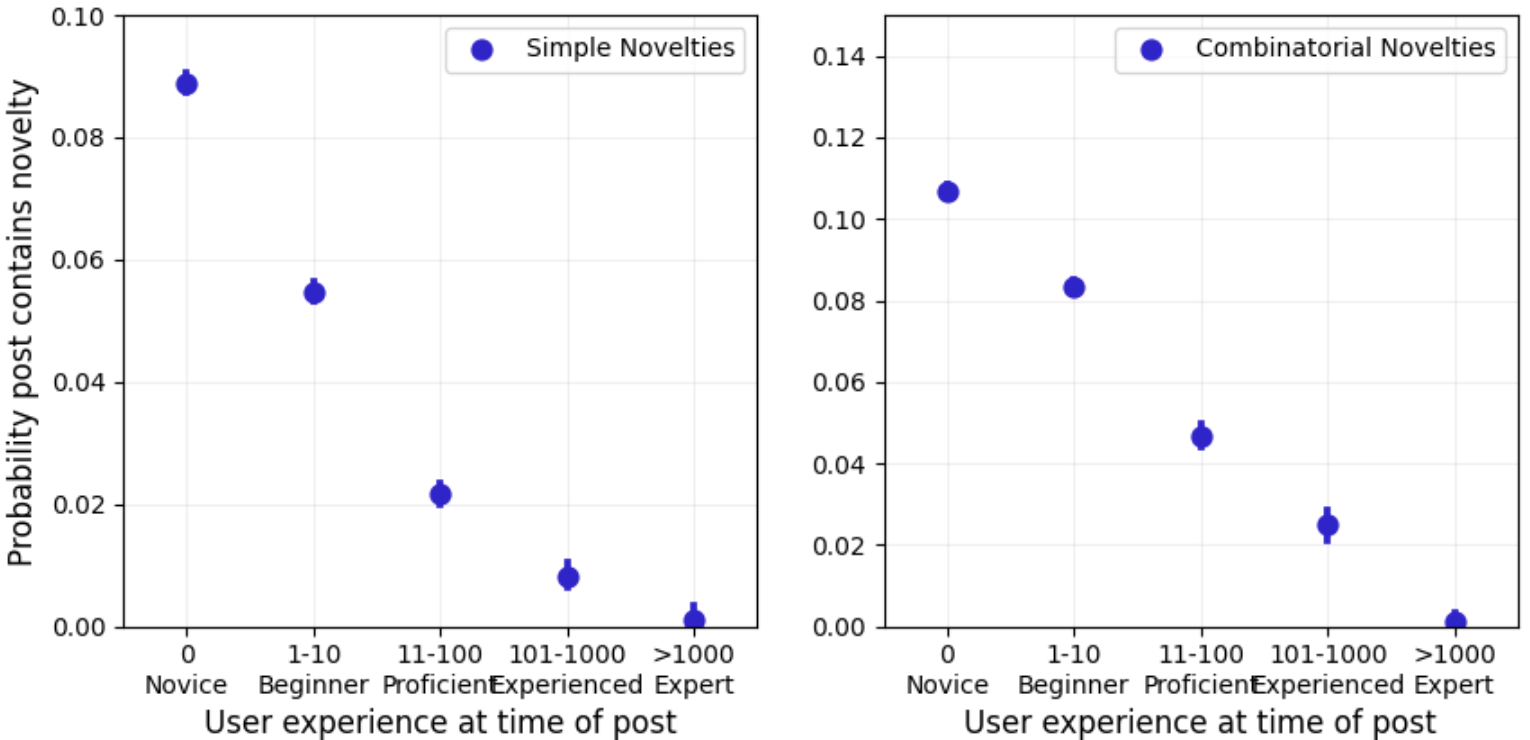}
    \caption{User Experience Distribution for Simple and Combinatorial Novelties among libraries appearing at least 100 times - restricted view to 2016.}
    \label{fig:example-user_exp_pairs_limit_100_year_2016}
\end{figure*}
\begin{figure*}
    \centering
    \includegraphics[width=0.8\textwidth]{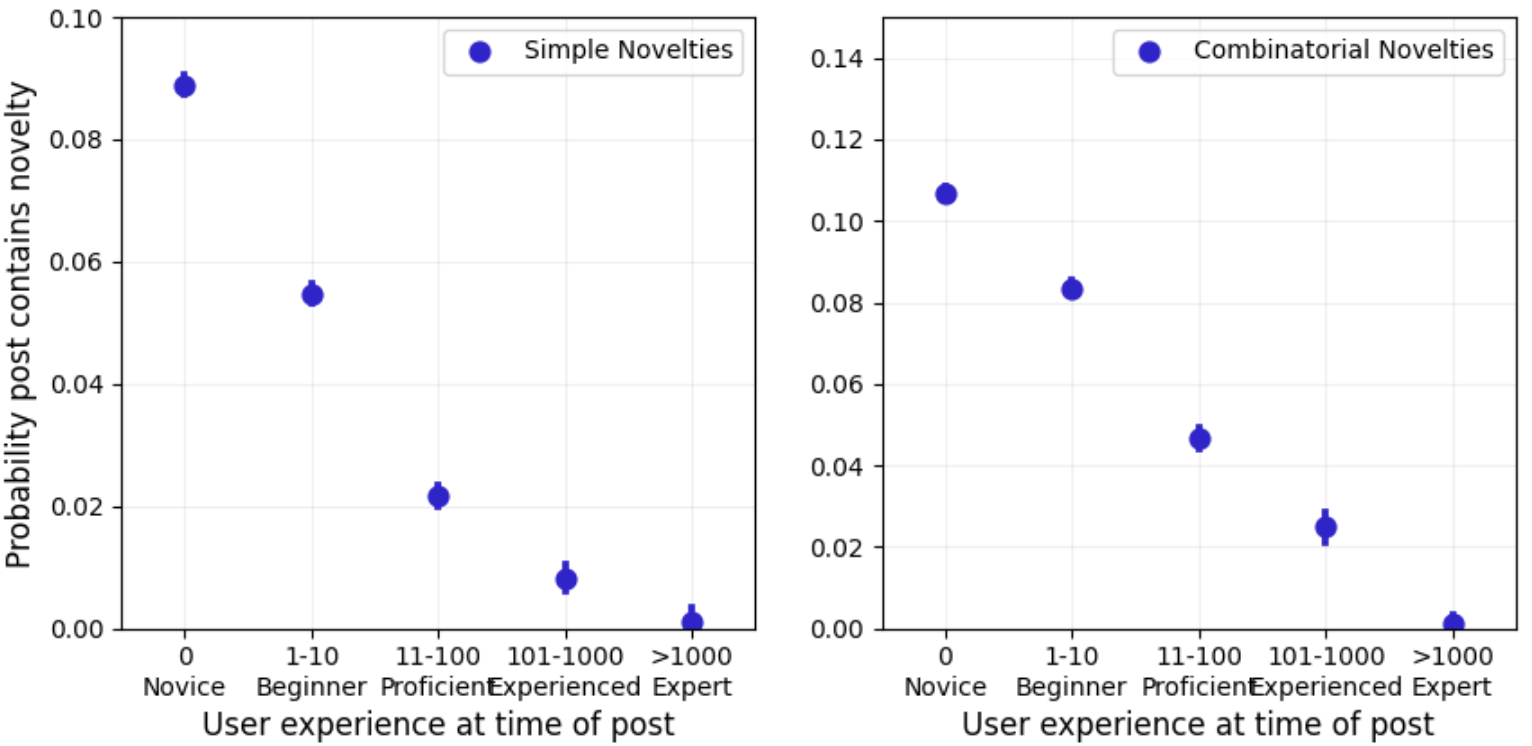}
    \caption{User Experience Distribution for Simple and Combinatorial Novelties among libraries appearing at least 1000 times - restricted view to 2016.}
    \label{fig:example-user_exp_pairs_limit_1000_year_2016}
\end{figure*}
\end{document}